\documentclass[%
 amsmath,amssymb,
 aps,
 twocolumn,
 showpacs,
 superscriptaddress,
 groupedaddress,
 prb
]{revtex4}

\usepackage{amsmath}
\usepackage{amssymb}

\usepackage{epsfig}

\usepackage{multirow}
\usepackage{diagbox}
\usepackage{graphicx}
\usepackage{dcolumn}
\usepackage{bm}
\usepackage{hyperref}
\usepackage[mathlines]{lineno}
\usepackage{xcolor}
\usepackage[english]{babel}
\makeatletter\AtBeginDocument{\let\@elt\relax}\makeatother

\def\<{\langle}
\def\>{\rangle}

\def\nn{\nonumber}

\def\beq{\begin{equation}}
\def\eeq{\end{equation}}

\newcommand{\bea}{\begin{eqnarray}}
\newcommand{\eea}{\end{eqnarray}}

\newcommand{\tdos}{$\mathrm{TDOS~}$}
\newcommand{\lut}{$\mathrm{LL~}$}
\newcommand{\luts}{$\mathrm{LLs~}$}
\newcommand{\rg}{$\mathrm{RG~}$}

\newcommand{\qh}{$\mathrm{QH~}$}

\newcommand{\tdosd}{$\mathrm{TDOS}$}
\newcommand{\tdosds}{$\mathrm{TDOSs~}$}
\newcommand{\lutd}{$\mathrm{LL}$}
\newcommand{\lutds}{$\mathrm{LLs}$}
\newcommand{\rgd}{$\mathrm{RG}$}
\newcommand{\qpcd}{$\mathrm{QPC}$}

\DeclareMathOperator{\sign}{sgn}

\def\lsim{\mathrel{\rlap{\lower4pt\hbox{\hskip1pt$\sim$}}
    \raise1pt\hbox{$<$}}}         
\def\gsim{\mathrel{\rlap{\lower4pt\hbox{\hskip1pt$\sim$}}
    \raise1pt\hbox{$>$}}}         

\usepackage{url}
\usepackage{amsfonts}
\usepackage{textcomp}
\def\BibTeX{{\rm B\kern-.05em{\sc i\kern-.025em b}\kern-.08em
    T\kern-.1667em\lower.7ex\hbox{E}\kern-.125emX}}

\begin{document}

\title{  Enhancement in tunneling density of states in 
  Luttinger liquid : \\  role of non-local interaction}

\author{Amulya Ratnakar  and Sourin Das \\
{\emph{Department of Physical Sciences,\\
Indian Institute of Science Education and Research (IISER) Kolkata \\
Mohanpur - 741 246,
West Bengal, India}}} 

\date{\today}
                              
\begin{abstract}

Power-law suppression of local electronic tunneling density of states (\tdosd) in the zero-energy limit is a hallmark of the Luttinger liquid (\lutd) phase of the interacting one-dimensional electron system. We present a theoretical model which hosts the \lut state with the surprising feature of enhancement rather than suppression in local \tdos originating from non local and repulsive density-density interactions. Importantly, we find enhancement of \tdos  in the manifold of parameter space where the system is stable in the renormalization group (\rgd) sense. We argue that enhancement of \tdos along with  \rg stability is possible only when the system has broken parity symmetry about the position of local \tdos enhancement. Such a model could be realized on the edge states of a bi layer quantum Hall system where both intra layer and inter layer density-density interactions are present mimicking the role of local and non local interactions, respectively.

\end{abstract}                              
                              
\maketitle

\section{Introduction}

It is well-known that the ampliude of electron tunneling into a Luttinger liquid (\lutd) state exhibits power-law suppression in the zero-energy limit owing to its non Fermi-liquid behavior~\cite{Kane1992,Yoreg1996,Fisher-Glazman, Eggert2000, kakashvili, Aristov2010, SDSarma2020, Fradkin,Fazio}. The suppression can be attributed to many-body orthogonality, which is akin to an orthogonality catastrophe discussed in the context of a \lut ~\cite{Anderson, Ogawa, Chen1992, gogolin1992, prokofev, Glazman_kane, Affleck_ludwig, Finkelstein1996, Furusaki}, and it can be understood as follows. When an electron-like quasiparticle (with vanishingly small energy)  tunnels locally into the \lut prepared in its ground state, interelectron interactions lead to a significant rearrangement of all the other electrons constituting the \lut state, which results in an excited state which is orthogonal to the corresponding ground state hence leading to the suppression of the tunneling process itself. This is a direct consequence of the fact that the low-energy excitation spectrum of a LL is devoid of electron like quasiparticles~\cite{SRao, Haldane, JVDelft, Maslov, Giamarchi, gogolinbook,  Meden_voit}. 

It is worthwhile to explore possibilities of departure from the observed suppression, which is treated as a hallmark of the \lut phase, and this is the main idea behind the present study. Here, we obtain an enhancement of local tunneling density of states (\tdosd) in the \lut phase.  We present a minimal set up that allows for such an enhancement in \tdos of a \lutd. It should be noted that in the bulk of a \lutd, any deviation from suppression is unlikely to take place due to the above-stated argument of the orthogonality. However,  in the local neighborhood of the boundary between two LLs, we may be able to realize a situation where such deviations could occur.  Hence, with the goal of finding a deviation from suppression, in this paper,  we consider a geometry involving the junction of two chiral \lutds ~\cite{Nancy, Chklovskii1998,Sen2008}.  
The junction of multiple \lutds~\cite{Nayak1999,Altland_gefen,chamon2003,Chamon2008,SDas2004,
SDas2006,SDas2008,Das2008,SDas2009,SDas2010,shi2016,
Agarwal2015,SLal2002,Egger2003,Sedeki,Demler2008,
Schonhammer2009,Rahmani,Aristov2011,Meden2000,Bellazzini2009,
Guo2006,Manu2016,Manu2020,Ramos2019,Aristov2013,chang2003,Wen}, whether chiral or non chiral, has been an area of theoretical interest owing to the rich physics associated with various fixed points that they host and belongs to the realm of the general topic of quantum impurity problems in low-dimensional electronic systems.

In an earlier study~\cite{SDas2009} involving one of the present authors along with others, it was shown that the exotic fixed points of a three-wire junction can lead to an
enhancement of electron \tdos in the vicinity of the junction when the \lut parameter $K$ is tuned to the repulsive interelectron interaction limit, i.e., $K<1$.
The origin of this \tdos enhancement was attributed to the reflection of a hole current~\cite{chamon2003} from the three-wire \lut junction due to interaction effects. A concern that remained is that the fixed points which allowed for an enhancement were unstable to relevant perturbation in the renormalization group (\rgd) sense and may not be direct of interest for experimental exploration. In the follow-up work,  the spin degree of freedom was incorporated for the three-wire \lut junction~\cite{Agarwal2015}; however, the issue of stability remained. Recently, a density-matrix renormalization group (DMRG) study was carried out by one of the present authors along with others which demonstrated a \tdos enhancement for the three-wire junction but again for an unstable fixed point \cite{Manu2020}.


\begin{figure}[t!]
\centering
\includegraphics[scale = 0.20]{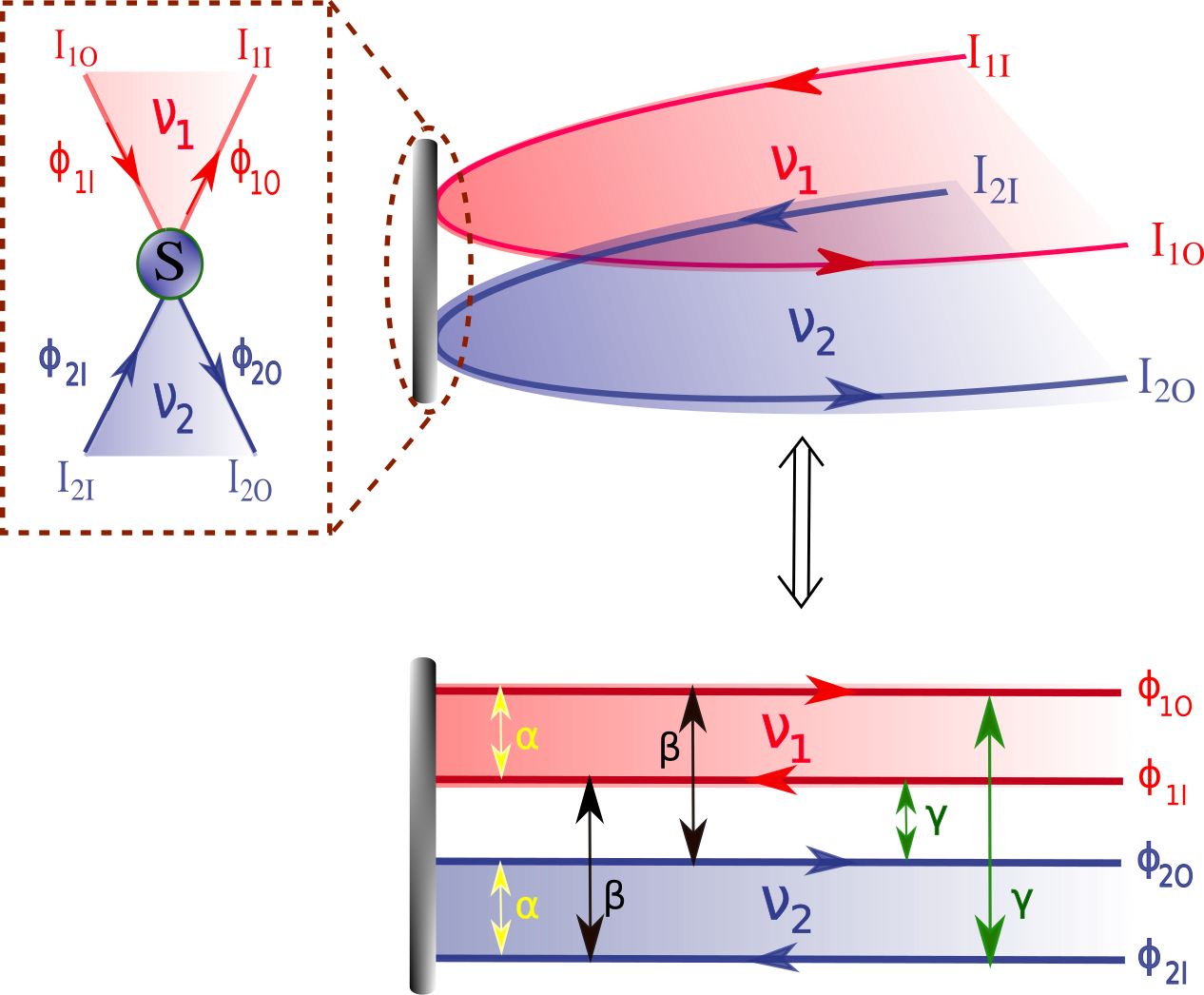}
\caption{Stacking of bilayer \qh states with filling fraction $\nu_{1}$ and $\nu_{2}$ exposed to a uniform magnetic field with a local tunnel coupling at the apex which is denoted by {\bf{S}}. The top left panel shows the zoomed-in view of the unfolded version of the bilayer \qh edge states. The bottom panel shows the $\alpha, \beta,$ and $\gamma$ interactions between chiral \qh edge states. Here, the subscript $``I"$ and $``O"$ stand for the fields flowing into the junction and fields the flowing out of the junction, respectively. $``I_{1/2,I/O}"$ stand for the currents on the incoming and outgoing edges.
}
\label{fig:bylayer_FQH}
\end{figure}


Actually, it is quite logical that all the fixed points depicting \tdos enhancement in the vicinity of the \lut junction are unstable fixed points. To understand
this point, let us consider a hypothetical situation consisting of a two-\lut wire junction tuned to a disconnected fixed point. The RG flow for the \tdos for each individual wire in the vicinity of the junction decides the rate at which the tunneling amplitude of an electron into the individual wires diverges or gets suppressed in the zero-energy limit ($E \rightarrow 0$). In the presence of a weak tunneling amplitude between the two wires, the net current flowing from one wire to another, in the linear response limit, is proportional to the product of the \tdosds of the two wires at the junction. In order for the disconnected fixed point to be stable against weak interwire electron tunneling perturbation, the tunneling current between the two wires must vanish in the zero-bias limit. This, in the RG sense, implies that the interwire tunneling operator is an irrelevant perturbation. Hence the stability of the disconnected fixed point along with simultaneous \tdos enhancement is achieved only when the rate at which \tdos gets suppressed in one of the wires is more than the rate at which \tdos is enhanced in the other wire as $E \rightarrow 0$, i.e., simultaneous \tdos enhancement and stability are always accompanied by breaking of parity symmetry between the two wires about the junction. This can be achieved by having different intrawire interactions in the two wires.

In this paper, we consider a fixed point of junction between two chiral \luts for exploring \tdos enhancement where the two chiral edge modes could belong to two distinct quantum Hall (QH) states. This is theoretically equivalent to having a single quantum point contact (\qpcd) in a Hall bar geometry forming a tunnel junction between the left and the right region, each of which in principle could host a distinct quantum Hall state surrounded by its own chiral edge states.  Such a setup is simple from a  theoretical perspective owing to the fact that a junction between two chiral \luts can host only two fixed points (a connected fixed point and the disconnected fixed point~\cite{kane_fisher,Wen}), unlike the three-wire case, which hosts a family of fixed points~\cite{chamon2003, SDas2006, SDas2009}.

Furthermore, we are interested in exploring how nonlocal density-density interaction between edge states belonging to the two sides of a QPC influences the \tdos. One way to simulate such interactions is to consider a situation where we fold the two-dimensional system about the QPC to form a bilayer system \cite{West1990,Eisentein,Shayegan,Eisentein1992} and then consider local (in the folded 1 D coordinate system) interlayer and intralayer edge interaction as shown in Fig.~\ref{fig:bylayer_FQH}. As discussed above, to host a fixed point that is stable and at the same time also shows \tdos enhancement, we need to break the junction symmetry.  We explore two different ways of breaking the symmetry (i) by having a junction between the chiral \luts belonging to two layers such that each layer has a different filling fraction or, (ii) if the filling fraction is same, then by introducing asymmetric intralayer edge interaction.

\section{Interacting QH Edge Hamiltonian}

Consider the situation of a bilayer interacting \qh system with filling fractions $\nu_{1}$ and $\nu_{2}$ on the two layers as depicted  in Fig.~\ref{fig:bylayer_FQH}. To begin with, we consider repulsive density-density interactions between the edges such that it poses a symmetric situation about the junction and is parametrized by $\alpha$, $\beta,$ and $\gamma$, where $\alpha$ is the interaction between the counterpropagating edge states in the same \qh states (intralayer interaction), $\beta$ is the interaction between the copropagating edge states of the different \qh states (interlayer interaction), and $\gamma$ is the interaction between the counterpropagating edge states of the different \qh states (interlayer interaction).

The Hamiltonian for \qh edge states can be described in terms of bosonic fields. The fermionic field $\psi_{I/O}$ for the electron on the edge can be expressed in terms of the bosonic fields $\phi_{I/O}$ using the standard bosonization formula~\cite{Wen,SRao,Haldane,JVDelft,shankar1995,Maslov,Giamarchi} as $\psi_{I/O} \sim F_{I/O} \exp(i \phi_{I/O}/\nu)$, where subscript $I(O)$ describes in (out) fields. Here, ``in"(``out") is used to index the chiral fields which flow into the junction (out of the junction). $F_{I/O}$ are the corresponding Klein factors for in/out fields.  Then the bosonized interacting edge Hamiltonian describing our setup is given by

\begin{widetext}
\begin{eqnarray}
H &=& \hbar \pi v_{F} \int_{0}^{\infty} dx \left[ \left(\frac{\rho_{1 I}^{2} + \rho_{1 O}^{2}}{\nu_{1}}\right) + \left(\frac{\rho_{2 I}^{2} + \rho_{2 O}^{2}}{\nu_{2}}\right)
\right. 
 + \left. 2\alpha \left( \frac{\rho_{1I}\rho_{1O}}{\nu_{1}} + \frac{\rho_{2I}\rho_{2O}}{\nu_{2}}\right) \right. \nonumber \\
&&  \quad \quad \quad \quad  \quad \quad \quad \quad + \left. \frac{2\beta}{\sqrt{\nu_{1} \nu_{2}}} \left( \rho_{1I}\rho_{2I} + \rho_{1O}\rho_{2O} \right) \right. 
 + \left. \frac{2\gamma}{\sqrt{\nu_{1} \nu_{2}}} \left( \rho_{1I}\rho_{2O} + \rho_{1O}\rho_{2I} \right) \right],
\label{Eq:Main_interacting Hamiltonian}
\end{eqnarray}
\end{widetext}
where $\rho_{i,I/O} = \pm (1/2\pi)\partial_{x}\phi_{i,I/O}$ and they represent the electronic density operator for the in/out bosonic fields corresponding to filling fraction $\nu_{i}$ of the $i$th \qh layer ($ i \, \in \, \lbrace 1,2\rbrace$). $v_{F}$ is the Fermi velocity, which has been taken to be the same on all the edges. Note that the interaction parameters $\alpha$, $\beta$, and $\gamma$ are scaled appropriately in the above Hamiltonian so that the transformation which diagonalizes the above Hamiltonian stays algebraically simple. We use the folded basis to describe the junction such that all the \qh edge states lie between $x=0$ and $x = \infty$ with the junction  positioned at $x=0$. We applied the appropriate fixed-point boundary condition on the ``in" and the ``out" fields at the junction. The interacting Hamiltonian given in Eq.~(\ref{Eq:Main_interacting Hamiltonian}) along with the boundary condition describes the total system. In what follows we will closely follow the diagonalization procedure for the above Hamiltonian as was done in Ref.~[\onlinecite{DSen2009}]. To begin with, we can rewrite Eq.~(\ref{Eq:Main_interacting Hamiltonian})  in a compact form as

\begin{equation}
H = \frac{ \hbar v_{F} }{4 \pi} \int_{0}^{\infty} dx\:\:\nabla \bar{\phi}^{T}(x) \:\;K \:\;\nabla \bar{\phi}(x),
\label{Eq:interacting Hamiltonian}
\end{equation}

where the matrix $K$ is given by 

\begin{equation}
K = \begin{pmatrix}
              1 & \beta & -\alpha & -\gamma \\
              \beta & 1 & -\gamma & -\alpha \\
              -\alpha & -\gamma & 1 & \beta \\
              -\gamma & -\alpha & \beta & 1
              \end{pmatrix}
\end{equation}

and is written in the basis $\bar{\Phi}(x,t)$, which is given by 
\begin{equation}
(\Bar{\phi}_{1},\Bar{\phi}_{2},\Bar{\phi}_{3},\Bar{\phi}_{4})_{(x,t)} = \left(\frac{\phi_{1 O}}{\sqrt{\nu_{1}}},\;\frac{\phi_{2 O}}{\sqrt{\nu_{2}}},\;\frac{\phi_{1I}}{\sqrt{\nu_{1}}},\;\frac{\phi_{2I}}{\sqrt{\nu_{2}}}\right)_{(x,t)}.
\label{Eq:field relation}
\end{equation}

Then at $t = 0$, the mode decomposition for the field $\bar{\phi}_{a}$ is given by 
\begin{equation}
\bar{\phi}_{a }(x) = \int^{\infty}_{0} \frac{dk}{k}\left[ \bar{c}_{a,k}e^{i \epsilon_{a} k x} + \bar{c}^{\dagger}_{a,k}e^{-i\epsilon_{a} k x}\right],
\end{equation}
where $a \; \in \lbrace 1,2,3,4\rbrace$, with $\epsilon_{a} = +1$ for $a = \lbrace 1,2 \rbrace$ (for the outgoing field) and $\epsilon_{a} = -1$ for $a = \lbrace 3,4 \rbrace$ (incoming field). The commutation relation  for the bosonic annihilation and creation operator is given by  $\left[\bar{c}_{a,k},\bar{c}^{\dagger}_{b,k'}\right] = \delta_{ab} k \delta(k-k')$, which is consistent with the commutation relation of the bosonic field in the real-space basis $\bar{\phi}(x)$ given by $\left[ \Bar{\phi}_{a}(x), \Bar{\phi}_{b}(y)\right] =  i\pi\epsilon_{a} \delta_{ab}\sign (x-y)$. Since the relation between  original interacting fields $\phi_{i,I/O}$ and the transformed field $\bar{\phi}_{a}$ is given by Eq.~(\ref{Eq:field relation}), the annihilation  operators $c_{i I/O k}$ of the $\phi_{i I/O}$ field and  $\bar{c}_{a k}$ of the $\bar{\phi}_{a}$ field are  also related as $(c_{1 O k},c_{2 O k},c_{1 I k},c_{2 I k}) = (\sqrt{\nu_{1}}\;\bar{c}_{1 k},\sqrt{\nu_{2}}\;\bar{c}_{2 k},\sqrt{\nu_{1}}\;\bar{c}_{3 k},\sqrt{\nu_{2}}\;\bar{c}_{4 k})$, where $\nu_{i}$ is the filling fraction of the $i$th QH layer.

Let the interacting $\bar{\phi}(x,t)$ field be related to the Bogoliubov field $\tilde{\phi}(x,t)$ through a real matrix $X$, such that
\begin{equation}
\bar{\phi}(x,t) = X \tilde{\phi}(x,t),
\label{Eq: phi_bar=X phi_tilde}
\end{equation} 
where
\begin{equation}
\tilde{\phi}_{\alpha}(x,t) = \int_{0}^{\infty}\frac{d k}{k}\left( \tilde{c}_{\alpha,k}e^{i \epsilon_{\alpha} k (x - \tilde{v}_{\alpha} t)} +  \tilde{c}^{\dagger}_{\alpha,k}e^{-i \epsilon_{\alpha} k (x - \tilde{v}_{\alpha} t)}\right),
\label{Eq: bogoliubov_decompose}
\end{equation}

where $\alpha \in \lbrace 1,2,3,4\rbrace$ and  $\epsilon_{\alpha} = \sign(\tilde{v}_{\alpha})$. $\tilde{c}_{\alpha k}$ ($\tilde{c}^{\dagger}_{\alpha k}$) is the bosonic  annihilation (creation) operator for the $\alpha$th Bogoliubov mode, with commutation relations as $\left[ \tilde{c}_{\alpha k }, \tilde{c}^{\dagger}_{\beta k'} \right] = \delta_{\alpha \beta}\; k\; \delta(k-k')$ and $\left[ \tilde{c}_{\alpha k }, \tilde{c}_{\beta k'} \right] = 0$, and this is consistent with  $\left[ \tilde{\phi}_{\alpha }(x) , \tilde{\phi}_{\beta}(y) \right] = \pi i\epsilon_{\alpha}\delta_{\alpha \beta}\sign(x-y)$.

From Eq.~(\ref{Eq:interacting Hamiltonian}), the Heisenberg equation of motion for the bosonic fields is given by  
\begin{equation}
\frac{d}{dt}\bar{\phi}_{a}(x,t) = -v_{f}\; \epsilon_{a}\sum_{\alpha=1}^{4} K_{a \alpha}\frac{d}{dx}\bar{\phi}_{\alpha}(x,t) ~.
\label{Eq: Eqn of motion}
\end{equation}

Using Eqs.~(\ref{Eq: phi_bar=X phi_tilde}) and (\ref{Eq: bogoliubov_decompose}) in Eq.~(\ref{Eq: Eqn of motion}), we have

\begin{widetext}
\begin{eqnarray}
\sum_{\alpha=1}^{4}\left[\int_{0}^{\infty} \frac{dk}{k} (i k \epsilon_{\alpha})\left( X_{a \alpha} \tilde{v}_{\alpha} - v_{f} \epsilon_{a} \sum_{b=1}^{4} K_{a b}X_{b \alpha} \right)\left( \tilde{c}_{\alpha k} e^{i k (x-\tilde{v}_{\alpha} t)} - \tilde{c}^{\dagger}_{\alpha k} e^{-i k (x-\tilde{v}_{\alpha} t)} \right)\right] = 0~.
\label{Eq: Diagonalization condition}
\end{eqnarray}
\end{widetext}
Equation~(\ref{Eq: Diagonalization condition}) implies that 
\begin{equation}
v_{f}\sum_{b=1}^{4} \epsilon_{a} K_{a b}X_{b \alpha} = X_{a \alpha} \tilde{v}_{\alpha}~.
\end{equation}

Now we can solve for the $X_{a\alpha}$ and the  $\tilde{v}_{\alpha}$ by solving the above equation. The $\tilde{v}_{\alpha}$'s are given by $\pm \sqrt{(1-\beta)^{2} - (\alpha-\gamma)^{2}}$ and $ \pm \sqrt{(1+\beta)^{2} - (\alpha + \gamma)^{2}}$, with a $+$ $(-)$ sign for out (in) free field. The bosonic excitations are stable if the $\tilde{v}_{\alpha}$'s are real. In order for new fields  to satisfy the bosonic commutation relations we must impose the following  normalized condition:
\begin{eqnarray}
\sum_{\alpha=1}^{4} \epsilon_{a} \epsilon_{\alpha} X_{a\alpha} X_{b \alpha} &=& \delta_{a b}~, \nonumber\\
\sum_{a=1}^{4} \epsilon_{a} \epsilon_{\alpha} X_{a\alpha} X_{a \beta} &=& \delta_{\alpha \beta}~.
\end{eqnarray}
Once we have obtained the $X$ matrix, then for all $(x,t)$ we have 
\begin{eqnarray}
\bar{\phi}_{a}(x,t) &=& \sum_{\alpha = 1}^{4} X_{a\alpha}\tilde{\phi}_{\alpha}(x,t)~, \nonumber\\
\tilde{\phi}_{\alpha}(x,t) &=& \sum_{a = 1}^{4} \epsilon_{a}\epsilon_{\alpha} X_{a\alpha}\bar{\phi}_{a}(x,t)~.
\end{eqnarray}

The interacting bosonic field operator $\bar{c}_{a k}$ and the Bogoliubov field operator $\tilde{c}_{\alpha k}$ are related as

\begin{widetext}
\begin{eqnarray}
\bar{c}_{a k}  &=& \sum_{\alpha = 1}^{4} X_{a\alpha}\left( P_{a\alpha,+} \tilde{c}_{\alpha k } e^{i(\epsilon_{\alpha} - \epsilon_{a})kx} + P_{a\alpha,-} \tilde{c}^{\dagger}_{\alpha k } e^{-i(\epsilon_{\alpha} + \epsilon_{a})kx} \right)~, \nonumber \\
\tilde{c}_{\alpha k}  &=& \sum_{a = 1}^{4} X_{a\alpha}\left( P_{a\alpha,+} \bar{c}_{a k } e^{i(\epsilon_{a} - \epsilon_{\alpha})kx} - P_{a\alpha,-} \bar{c}^{\dagger}_{a k} e^{-i(\epsilon_{a} + \epsilon_{\alpha})kx} \right) ~,
\end{eqnarray}
\end{widetext}

where the projection operator  is given by $P_{a\alpha, \pm} = {(1 \pm \epsilon_{a} \epsilon_{\alpha})}/{2}$.
Let $\bar{\phi}_{O/I}$, $\tilde{\phi}_{O/I}$ be doublets  such that $\bar{\phi}_{O} = (\bar{\phi}_{1},\bar{\phi}_{2})$, $\bar{\phi}_{I} = (\bar{\phi}_{3},\bar{\phi}_{4})$ and $\tilde{\phi}_{O} = (\tilde{\phi}_{1},\tilde{\phi}_{2})$, $\tilde{\phi}_{I} = (\tilde{\phi}_{3},\tilde{\phi}_{4})$. Also, $\tilde{v}_{1} = -\tilde{v}_{3} > 0$ and $\tilde{v}_{2} = -\tilde{v}_{4} > 0$. Then, we can express Eq.~(\ref{Eq: phi_bar=X phi_tilde}) as  
\begin{equation}
\begin{pmatrix}
\bar{\phi}_{O} \\
\bar{\phi}_{I}
\end{pmatrix}_{(x,t)} = \begin{pmatrix}
                X_{1} & X_{2} \\
                X_{3} & X_{4}
                \end{pmatrix} \begin{pmatrix}
                              \tilde{\phi}_{O} \\
                              \tilde{\phi}_{I}
                              \end{pmatrix}_{(x,t)}~,
\label{Eq: couplet bar_tilde fields rlation}
\end{equation}

where the $X_{i}$'s are  $2\times 2$ matrices. Now, the original incoming fields $\phi_{i I}$ and outgoing field $\phi_{i O}$ are related to each other through a boundary condition at the junction $(x=0)$. The boundary condition is expressed as the current splitting matrix $S$, which corresponds to the different fixed points of the junction, such that
\begin{eqnarray}
\begin{pmatrix}
\phi_{1O} \\
\phi_{2O}
\end{pmatrix}_{(x=0,t)} &=& S \begin{pmatrix}
                              \phi_{1I} \\
                              \phi_{2I}
                              \end{pmatrix}_{(x=0,t)}~, \nonumber\\
\begin{pmatrix}
\bar{\phi}_{1} \\
\bar{\phi}_{2}
\end{pmatrix}_{(x=0,t)} &=& M^{-1}S M \begin{pmatrix}
                              \bar{\phi}_{3} \\
                              \bar{\phi}_{4}
                              \end{pmatrix}_{(x=0,t)} = \bar{S}\begin{pmatrix}
                              \bar{\phi}_{3} \\
                              \bar{\phi}_{4}
                              \end{pmatrix}_{(x=0,t)}~,\nonumber\\
\label{Eq: Field splitting matrix}
\end{eqnarray}

where $M$ is a $2 \times 2$ matrix, with $M_{ij} = \sqrt{\nu_{i}} \delta_{ij}$ and $\bar{S} = M^{-1} S M$. From Eqs.~(\ref{Eq: couplet bar_tilde fields rlation}) and (\ref{Eq: Field splitting matrix}), we have 

\begin{eqnarray}
\left( X_{1}\; \tilde{\phi}_{O} + X_{2}\; \tilde{\phi}_{I} \right)_{(x=0,t)} &=& \bar{S}\left( X_{3}\; \tilde{\phi}_{O} + X_{4}\;\tilde{\phi}_{I} \right)_{(x=0,t)}, \nonumber \\
\left( X_{1} - \bar{S} X_{3} \right)\tilde{\phi}_{O \; (x=0,t)} &=& \left( \bar{S} X_{4} - X_{2}\right)\tilde{\phi}_{I \; (x=0,t)},\nonumber\\
\end{eqnarray}
\begin{equation}
\tilde{\phi}_{O \; (x=0,t)} = \left( X_{1} - \bar{S} X_{3} \right)^{-1}\left( \bar{S} X_{4} - X_{2}\right)\tilde{\phi}_{I \; (x=0,t)} , 
\label{Eq: bogo_scattering_mat_relation}
\end{equation}

which can be translated to finite values of $x$ using the following relation,
\begin{equation}
\tilde{\phi}_{O} (x,t) = \left( X_{1} - \bar{S} X_{3} \right)^{-1}\left( \bar{S} X_{4} - X_{2}\right)\tilde{\phi}_{I} (-x,t).
\end{equation}
Here, we have used the fact that in our setup the incoming fields are left-moving fields (see Fig.~\ref{fig:bylayer_FQH}). Now, using Eq.~(\ref{Eq: couplet bar_tilde fields rlation}) and the relation between the $\bar{\phi}_{a}$ and $\phi_{a}$ fields, we have
\begin{eqnarray}
\phi_{O}(x,t) && = M\left[T_{1} \tilde{\phi}_{I}(-x,t) + T_{2}\tilde{\phi}_{I}(x,t)\right], \nonumber\\
\phi_{I}(x,t) &&= M\left[T_{3} \tilde{\phi}_{I}(-x,t) + T_{4}\tilde{\phi}_{I}(x,t)\right] ,
\label{modified_bosonic_field} 
\end{eqnarray}
where,
\begin{eqnarray}
T_{1} &=& X_{1}\left(X_{1} - \bar{S}X_{3}\right)^{-1}\left(\bar{S}X_{4}-X_{2}\right), \nonumber\\
T_{2} &=& X_{2}, \nonumber\\
T_{3} &=& X_{3}\left(X_{1} - \bar{S}X_{3}\right)^{-1}\left(\bar{S}X_{4}-X_{2}\right), \nonumber\\
T_{4} &=& X_{4} .
\end{eqnarray}
Hence we have expressed all the interacting bosonic fields in terms of the tilde fields [Eq.~(\ref{modified_bosonic_field})], which are free, and this will be used to calculate TDOS and scaling dimensions of tunneling and backscattering operators that could be switched on at the junction for RG analysis. Before we conclude this section, it should be noted that for the setup considered here, there are only two allowed fixed points \cite{Wen,Sen2008} and, hence, two possible $S$ matrices, which are given by
\begin{equation}
S_{1}=\begin{pmatrix}
      1 & 0 \\
      0 & 1 
      \end{pmatrix},
\label{Eq:S1 FP}
\end{equation}
\begin{equation}
S_{2} = \frac{1}{\nu_{1}+\nu_{2}}\begin{pmatrix}
                                   \nu_{1} - \nu_{2} & 2\nu_{1} \\
                                   2\nu_{2} & \nu_{2} - \nu_{1}
                                   \end{pmatrix},
\label{Eq:S2 FP}
\end{equation}
where $S_{1}$ is the fully reflecting disconnected fixed point and $S_{2}$ is the strongly coupled fixed point. For $\nu_{1} \neq \nu_{2}$, the $S_{2}$ fixed point may allow for incident current to be partially reflected as a hole current.
%
\section{Power law dependence of \tdosd}
The electronic  \tdosd~\cite{kane_fisher, SDas2009} at energy $E$ at the position $x$ is given by

\begin{equation}
\rho(E) = 2\pi \sum_{n} \vert {_{N+1}\langle n |} \psi^{\dagger}(x)|0\rangle_{N} \vert^{2}\delta(E_{n}^{N+1} - E_{0}^{N} -E) ,\nonumber
\end{equation}    
where $E_{n}^{N+1}, \vert n \rangle_{N+1}$ and $E_{0}^{N}, \vert 0 \rangle_{N}$ are the energy eigenvalues and eigenstates corresponding to the $n$th excited state of $(N+1)$-electron system and the ground state of the $N$-electron system, respectively, for the interacting Hamiltonian given in Eq.~(\ref{Eq:Main_interacting Hamiltonian}) subjected to appropriate boundary conditions ($S_1$ or $S_2$)and $\psi^{\dagger}(x)$ is the electron creation operator at position $x$. In particular, we will be calculating  \tdosds only for the outgoing edge as they only carry interesting information about the fixed point to which the junction is tuned. Hence, to obtain the \tdos in terms of the bosonic field, we rewrite it as
\begin{eqnarray}
\rho_{i}(E) &=& \int^{\infty}_{-\infty} \langle 0\vert \psi_{iO}(x,t)\psi^{\dagger}_{iO}(x,0)\vert0\rangle e^{-iEt} dt \nonumber,
\end{eqnarray}
which in terms of the bosonic fields $\phi_{i O}$ reads as 
\begin{equation}
\rho_{{i}}(E) \sim \int_{-\infty}^{\infty} dt \langle 0\vert e^{i\frac{\phi_{iO}(x,t)}{\nu_{i}}}e^{-i\frac{\phi_{iO}(x,0)}{\nu_{i}}}\vert0 \rangle e^{-iEt} \nonumber.
\end{equation}

Here, we have suppressed the subscript representing the number of electrons in the ground state given by $\vert0 \rangle$ for notational convenience. Using  Eq.~(\ref{modified_bosonic_field}), we evaluate the above expression in two limits, (i) at the junction (x=0), and (ii) far from the junction $(x \rightarrow \infty)$. In both these limits, \tdos has a pure power-law dependence of the form of $E^{\left(\Delta_{i} -1\right)}$. The \tdos power law at the junction ($x=0$) is denoted by $\Delta^{0}_{i}$, and far from the junction, $x\rightarrow \infty$ is denoted by $\Delta^{\infty}_{i}$ (for details, see Appendix \ref{TDOS}). After a straightforward algebra, the \tdos exponent at the junction is found to be given by
\begin{equation}
\Delta^{0}_{i} = \frac{1}{\nu_{i}}\sum_{j=1}^{2}\left(\left[T_{1}\right]_{ij}+\left[T_{2}\right]_{ij}\right)^{2} ,
\end{equation}
while far from the junction it is given by
\begin{equation}
\Delta^{\infty}_{i} = \frac{1}{\nu_{i}}\sum_{j=1}^{2}\left(\left[T_{1}\right]^{2}_{ij}+\left[T_{2}\right]^{2}_{ij}\right).   
\end{equation}
\tdos in the zero-energy limit is enhanced when $\Delta_{i}-1<0$, is marginal when $\Delta_{i} = 1$, and is suppressed when $\Delta_{i}-1>0$. Our primary focus is to study $\Delta^{0}_{i}$, but before we go ahead, we briefly discuss $\Delta^{\infty}_{i}$. $\Delta^{\infty}$ does not depend on the fixed point that we impose at the junction but gets modified only by the bulk interaction between the edges, and it always corresponds to suppressed \tdos  irrespective of the interaction strength which is expected from standard \lut physics~\cite{SDas2009}. The explicit form for the \tdos exponent $\Delta^{\infty}_{i}$ corresponding to our model considered in Eq.~(\ref{Eq:Main_interacting Hamiltonian}) is given by
\begin{eqnarray}
\Delta_{i}^{\infty} =&& \frac{1}{2\nu_{i}} \left( \frac{1-\beta}{\sqrt{(1-\beta)^{2} - (\alpha - \gamma)^{2}}} \right. \nonumber \\
&& \left. + \frac{1+\beta}{\sqrt{(1+\beta)^{2} - (\alpha + \gamma)^{2}}} \right). 
\end{eqnarray}
Note that in the $\alpha,\beta,\gamma \rightarrow 0$ limit we recover the expected $1/\nu$  power-law suppression of \tdos for an edge of the fractional quantum Hall state~\cite{chang2003}. The power law of $1/\nu$  is also recovered when only  $\alpha, \gamma \rightarrow 0$ while  $\beta \neq 0$, due to the fact that the nonzero $\beta$ corresponds to a pure forward scattering interaction and hence can result only in the renormalization of Fermi velocity but cannot influence the power law of the \tdosd. One should also note that even in absence of the tunneling between the edges at $x=0$, the very presence of interaction parameters $\alpha, \gamma$ breaks translational invariance along the edge while $\beta$ alone  does not affect translational invariance as expected.

\section{Stability of the Fixed Point}

In this section, we obtain a general expression for the scaling dimension of various tunneling and backscattering operators which can be switched on at the junction, where the scaling dimension being greater (less) than unity corresponds to an irrelevant (relevant) operator and being equal to 1 corresponds to being marginal. There are two possible fixed points for the junction described in Fig.~\ref{fig:bylayer_FQH}: (i) The first one is the disconnected fixed point, where the tunneling between the two layers at $x=0$ is fully suppressed. Hence the most important perturbation to be analyzed as far as the \rg stability of the junction is concerned is the electron tunneling operator between the two layers at $x=0$. (ii) The second one is the strong tunneling fixed point, where the two layers are strongly coupled at $x=0$ and, hence, the most important perturbation to be analyzed as far as the \rg stability of the junction is concerned is the quasiparticle backscattering operator in each layer at $x=0$.

Furthermore, it should be noted that the relation between the scaling dimensions of tunneling operators, which can be switched on at the junction as a perturbation, and the \tdos in the immediate vicinity of the junction ($x\rightarrow 0$) is not a simple relation which one might naively expect. To understand this point, let us consider the disconnected fixed point to be specific. In this case, the scaling dimension of interlayer tunneling operators is dictated by the correlation function given by
$G=\langle 0\vert\psi^{\dagger}_{e,1 O}(0)\, \psi_{e,2 I}(0) \, \psi^{\dagger}_{e,2 I}(t)\, \psi_{e,1 O}(t)\vert 0\rangle$, while the TDOS in each of the individual edge states is governed by the correlation functions $g_1 = \langle 0\vert\psi^{\dagger}_{e,1O}(0,0)  \psi_{e,1 O}(0,t)\vert 0\rangle$ and $g_2 =\langle 0\vert\psi^{\dagger}_{e,2 I}(0,0)  \psi_{e,2 I}(0,t)\vert 0\rangle$.  Here, the subscript ``$e$" corresponds to the electron operator, while we will use ``$qp$" for the quasiparticle operator and the subscript $1,2$ stand for the layer index. Hence one might expect that $G = g_1 g_2$ for the disconnected fixed point leading to a simple relation between \tdos and the stability of the junction. However, $G \neq g_1 g_2$ owing to the fact that one has interlayer interactions such that the ground state of the full edge Hamiltonian, $\vert 0\rangle$, does not decompose onto the direct product of the ground states of the edge Hamiltonian of individual layers, i.e., $\vert 0\rangle \neq \vert 0\rangle_1 \vert 0\rangle_2$ even for the disconnected fixed point. This fact plays an important role in the interplay of stability of a fixed point and \tdos  enhancement via the various nonlocal interaction terms.

The expressions for scaling dimension of backscattering and tunneling operators are straightforward to calculate using Eq.~(\ref{modified_bosonic_field}) and are given below:

\begin{enumerate}
\item[(1)]
 The intralayer quasiparticle backscattering operator $\psi^{\dagger}_{qp,i O}(0)\psi_{qp,i I}(0)$ has a  scaling dimension given by $d_{ii}^{O,I}=d_{ii}^{I,O}  = \frac{1}{2}\sum_{k=1}^{2}(\Lambda_{B, i}^{k})^{2}$, where
\begin{equation}
\Lambda_{B,i}^{k} = \sqrt{\nu_{i}}(T_{3} + T_{4} - T_{1} - T_{2})_{ik},
\label{Eq: dii}
\end{equation}

\item[(2)] The interlayer electron tunneling operator $\psi^{\dagger}_{e,i O}(0)\psi_{e,j I}(0)$ has a scaling dimension given by $d_{ij}^{O,I} =d_{ji}^{I,O}= \frac{1}{2}\sum_{k=1}^{2}(\Lambda_{T, ji}^{k})^{2}$, where
\begin{equation}
\Lambda_{T,ij}^{k} = \frac{1}{\sqrt{\nu_{i}}} (T_{3} + T_{4})_{ik} - \frac{1}{\sqrt{\nu_{j}}} (T_{1} + T_{2})_{jk}, 
\label{Eq: dij}
\end{equation}

\item[(3)] The interlayer electron tunneling operator $\psi^{\dagger}_{e,i O}(0)\psi_{e,j O}(0)$ has a scaling dimension given by $d_{ij}^{O,O} = d_{ji}^{O,O} = \frac{1}{2}\sum_{k=1}^{2}(\Lambda_{T, ij}^{k})^{2}$, where
\begin{equation}
\Lambda_{T,ij}^{k} = \frac{1}{\sqrt{\nu_{i}}}(T_{1} + T_{2})_{ik} - \frac{1}{\sqrt{\nu_{i}}}(T_{1} + T_{2})_{jk} ,
\label{Eq: dOO}
\end{equation}

\item[(4)] The interlayer electron tunneling operator $\psi^{\dagger}_{e,i I}(0)\psi_{e,j I}(0)$ has a scaling dimension given by $d_{ij}^{I,I} = d_{ji}^{I,I} = \frac{1}{2}\sum_{k=1}^{2}(\Lambda_{T, ij}^{k})^{2}$, where
\begin{equation}
\Lambda_{T,ij}^{k} =  \frac{1}{\sqrt{\nu_{i}}} (T_{3} + T_{4})_{ik} -  \frac{1}{\sqrt{\nu_{i}}} (T_{3} + T_{4})_{jk}.
\label{Eq: dII}
\end{equation}

\end{enumerate}

\begin{figure*}[t]
\centering
\includegraphics[scale = 0.35]{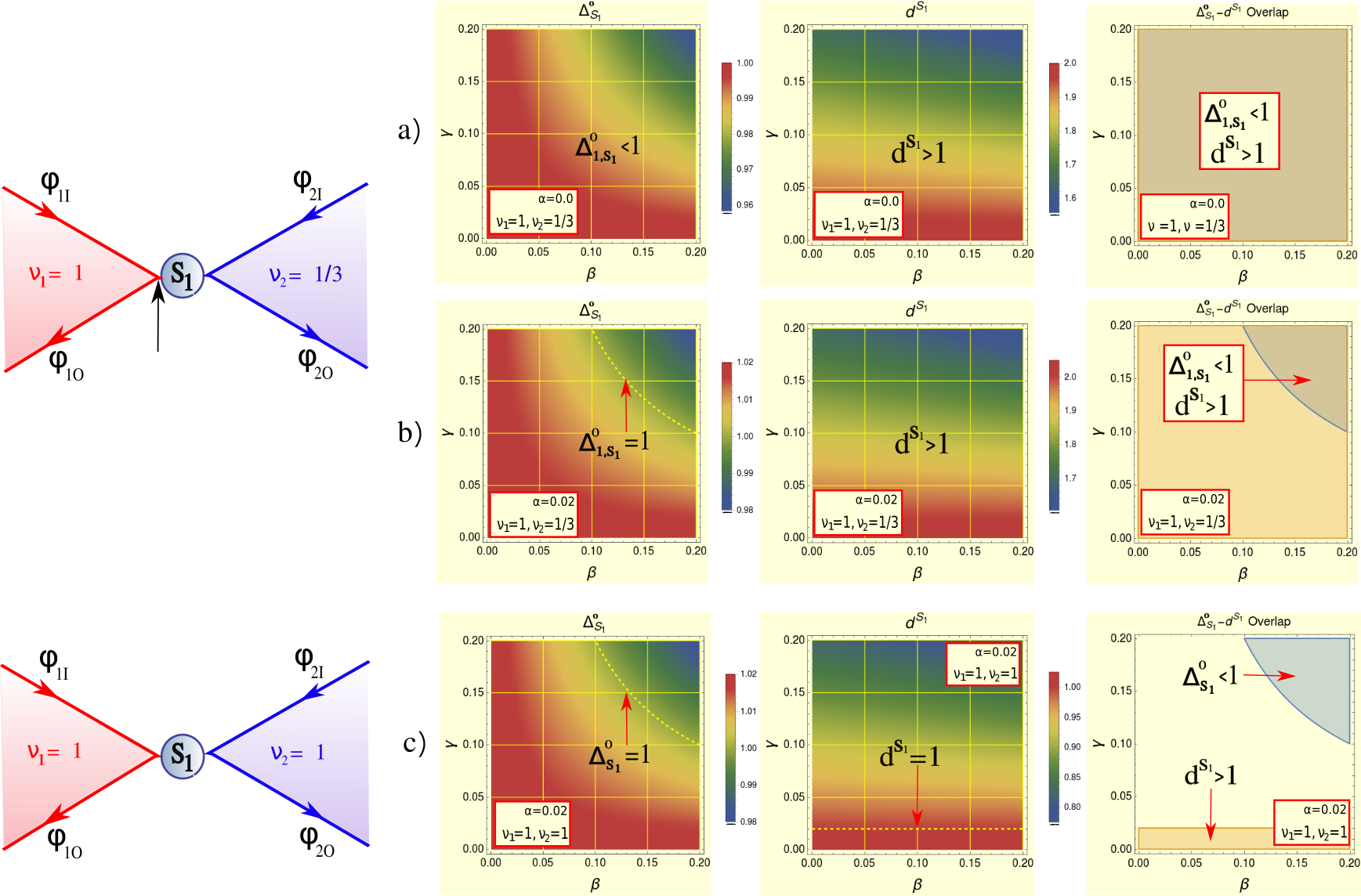}
\caption{The schematic pictures on the left show the unfolded version of the $S_{1}$ junction fixed point of bilayer \qh states exposed to a uniform magnetic field. For a junction of $\nu_{1} = 1$ and $\nu_{2} = 1/3$ tuned to the $S_{1}$ fixed point, (a) and (b) each show three density plots corresponding to $\Delta^{0}_{S_{1}}$ and $d^{S_{1}}$ and the region of simultaneous \tdos enhancement and stability as we move from left to right in each row for $\alpha = 0$ and $\alpha =0.02$ respectively. (c) shows $\Delta^{0}_{S_{1}}$ and $d^{S_{1}}$ and the region of simultaneous \tdos enhancement and stability as we move from left to right in the row
for $\alpha = 0.02$. The third plot in this row indicates that the region of simultaneous TDOS enhancement and stability is mutually exclusive in this case.
}
\label{fig:S1 FP plots-pert}
\end{figure*}


\begin{figure*}[t]
\centering
\includegraphics[scale = 0.35]{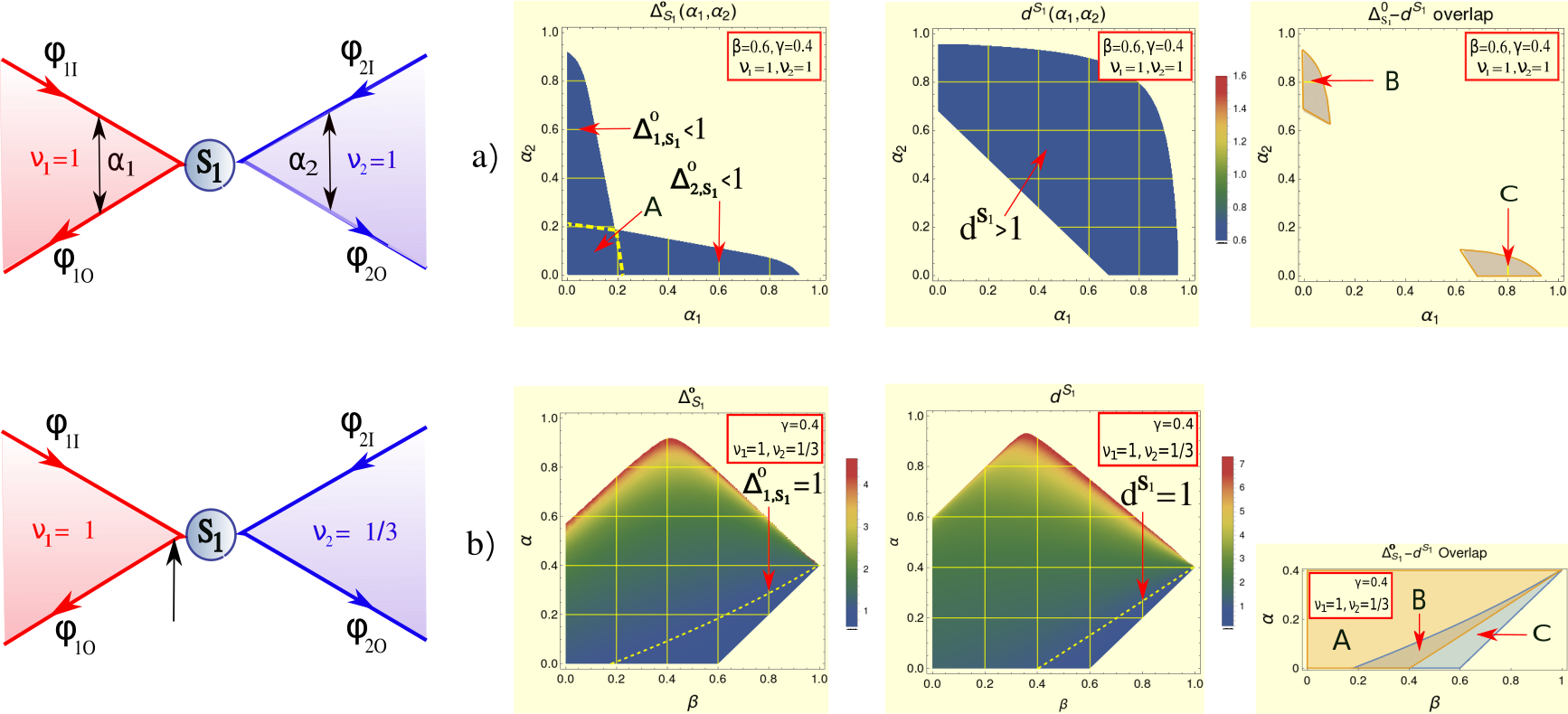}
\caption{The schematic pictures on the left show the unfolded version of the $S_{1}$ junction fixed point of bilayer \qh states exposed to a uniform magnetic field. For a junction of $\nu_{1} = 1$ and $\nu_{2} = 1$ states tuned to the $S_{1}$ fixed point with asymmetric $\alpha$ in two layers, (a) shows three density plots corresponding to $\Delta^{0}_{S_{1}}$ and $d^{S_{1}}$ and the region of simultaneous \tdos enhancement and stability as we move from left to right in the row for $\beta = 0.6$ and $\gamma = 0.4$. Here, region A shows the interaction parameters for which TDOS is enhanced on both the edges. Region B (C) shows interaction parameters for which \tdos for the $\nu_{1}$ ($\nu_{2}$) QH edge is enhanced and the junction is stable simultaneously. For a junction of $\nu_{1} = 1$ and $\nu_{2} = 1/3$ tuned to the $S_{1}$ fixed point, in the presence of symmetric interactions, (b) shows three density plots corresponding to $\Delta^{0}_{S_{1}}$ and $d^{S_{1}}$ and the region of simultaneous \tdos enhancement and stability as we move from left to right in the row for $\gamma = 0.4$.  Here, regions A, B, and C in the rightmost plots correspond to $(\Delta^{0}_{S_{1}} > 1,d^{S_{1}} > 1)$, $(\Delta^{0}_{S_{1}} < 1,d^{S_{1}} > 1)$, and $(\Delta^{0}_{S_{1}} < 1,d^{S_{1}} < 1)$, respectively. Region B shows the interaction parameters for which the TDOS is enhanced and the junction is stable simultaneously.
}
\label{fig:S1 FP plots}
\end{figure*}

\section{Simultaneous \tdos Enhancement and Stability of $S_{1}$ fixed point}

The explicit form of the \tdos exponent  corresponding to the disconnected fixed point $S_{1}$, denoted by $\Delta^{0}_{i,S_{1}}$, is evaluated on one of the two outgoing \qh edge states of the bi-layer system and is given by 
\begin{eqnarray}
\Delta_{i,S_{1}}^{0} =&& \frac{1}{2\nu_{i}} \left( \sqrt{\frac{1 + \alpha -\beta -\gamma}{1-\alpha - \beta + \gamma}} \,  \right. \nonumber \\
&& \left. + \sqrt{\frac{1+\alpha+\beta + \gamma}{1-\alpha + \beta -\gamma}} \right) .
\end{eqnarray}

We note that though interlayer interactions do exist, the \tdos in each layer only depends on the filling fraction ($\nu_{i}$) of the respective layer and not on that of the other layer. We will see later that this is not the case for the $S_{2}$ fixed point. It is also clear from the above expression that increasing $\alpha$ monotonically increases $\Delta_{i,S_{1}}^{0}$, which leads to the suppression of \tdosd. When $\beta$ and $\gamma$ are zero in the above expression, $\Delta_{i,S_{1}}^{0}$ reduces to  $(1/\nu_i)\sqrt{(1+\alpha)/(1-\alpha)}$, where  $\sqrt{(1+\alpha)/(1-\alpha)}$ is nothing but the inverse of the standard \lut parameter~\cite{Fisher1997} which is known to suppress the \tdos and the factor $1/\nu_i$ leads to additional suppression owing to the presence of a fractional \qh edge state \cite{chang2003}. Also, it was discussed earlier that the effect of $\beta$ alone is trivial as it represents the forward scattering interaction. Hence the enhancement of \tdos is expected to be induced by the presence of a finite $\gamma$.

To have a closer look at the interplay of various interaction  parameters leading to the enhancement of \tdos, we carry out a small $ \alpha,\gamma$ expansion of $\Delta_{i, S_{1}}^{0}(\alpha,\beta,\gamma)$ around $(\alpha=0,\beta,\gamma=0)$ to leading order and obtain
\begin{equation}
\Delta_{i, S_{1}}^{0} \simeq \frac{1}{\nu_{i}}\left( 1+\frac{\alpha - \beta \gamma}{1-\beta^{2}} \right) .
\label{pertur}
\end{equation}
From now onwards, we will only consider the case of repulsive electron-electron interactions, i.e., $\alpha,\beta,\gamma > 0$. Furthermore, we focus on a specific case for exploring the possibility of observing enhancement of \tdos (i.e., $\Delta^{0}_{i,S_{1}}<1$) for a junction of a $\nu_1=1$ and $\nu_2=1/3$ QH system. This case could be of relevance as in this case we break the layer symmetry (which is necessary  for the observation of simultaneous \tdos enhancement and stability of the junction) by choosing distinct $\nu$ for each layer and both $\nu=1$ and $\nu = 1/3$ represent a quantum Hall state which depicts prominent plateaus in experiments \cite{laughlin1981,stormer1999,Halperin1982}. It is expected that \tdos enhancement for the $\nu_2=1/3$ edge will be practically impossible due to strong suppression arising from the $1/\nu$ term in $\Delta_{i,S_{1}}^{0} $, and hence we focus on the $\nu_1=1$ edge only. Equation~(\ref{pertur}) implies that \tdos enhancement for the $\nu_1=1$ edge will be possible only if  $\beta \, \gamma > \alpha$ in the small $\alpha, \gamma$ limit, which implies that the magnitude of $\alpha,\beta$, and $\gamma$ has to follow a specific hierarchy for \tdos enhancement. However, most importantly, this inequality points to the fact that interaction parameters $\gamma$ and $\beta$ are essential for \tdos enhancement while $\alpha$ is not (i.e., $\alpha$ can be zero). This point is demonstrated numerically in Fig.~\ref{fig:S1 FP plots-pert}, where the first plot in  Fig.~\ref{fig:S1 FP plots-pert}(a) shows enhancement of the \tdos in the $\beta$-$\gamma$ plane around the origin whereas the first plot in Fig.~\ref{fig:S1 FP plots-pert}(b) shows that the region of \tdos enhancement starts shrinking as we turn on small but finite $\alpha$.

As far as the stability of the $S_{1}$ fixed point (FP) is concerned, the most relevant operators are the interlayer single electron tunneling operators, which are to be considered for the analysis because the scaling dimension of back-scattering operators is $d_{11}^{I,O}= d_{22}^{I,O}= 0$ for all $\alpha, \beta, \gamma$ as expected. Also note that for the $S_{1}$ FP, $d_{12}^{I,O}=d_{12}^{O,I} =d_{12}^{I,I}=d_{12}^{O,O} = d^{S_{1}}$. We obtain the expression for $ d^{S_{1}}$, which is given by
\begin{equation}
\begin{aligned}
d^{S_{1}} = {} & \frac{1}{4}\left( \sqrt{\frac{1+\alpha - \beta -\gamma}{1-\alpha - \beta + \gamma}}\left(\frac{1}{\sqrt{\nu_{i}}} + \frac{1}{\sqrt{\nu_{j}}}\right)^{2}\right. \\
& + \left. \sqrt{\frac{1+\alpha + \beta +\gamma}{1-\alpha + \beta - \gamma}}\left(\frac{1}{\sqrt{\nu_{i}}} -\frac{1}{\sqrt{\nu_{j}}} \right)^{2}\right).
\end{aligned}                
\end{equation}

Note that $d^{S_{1}}$ is a function of both symmetric and antisymmetric combination of $\sqrt{\nu_{1}}$ and $\sqrt{\nu_{2}}$. The presence of antisymmetric combination indicates that the broken layer symmetry ($\nu_{1} \neq \nu_{2}$) results in an additional contribution to the scaling dimension which is connected to the essential requirement for tuning simultaneous \tdos enhancement and stability. It is also clear from the above expression that the junction gets more and more stable as we increase $\alpha$; that is, increasing $\alpha$ leads to monotonically increasing $d^{S_{1}}$. Hence finite $\alpha$ has an adverse effect on simultaneous \tdos enhancement and stability as its presence, on one hand, leads to greater stability but, on the other hand, suppresses the enhancement of \tdos.

Similar to the expansion of $\Delta_{i, S_{1}}^{0}$ above, we now perturbatively expand $d^{S_{1}}$ about $(\alpha=0,\beta,\gamma=0)$ to obtain the following expression:

\begin{eqnarray}
d^{S_{1}} &\simeq& \frac{\nu_{1} + \nu_{2}}{2\,\nu_{1} \nu_{2}}\left(1+\frac{\alpha - \beta \gamma}{1-\beta^{2}}+
\left(\frac{2\sqrt{\nu_{1}\nu_{2}}}{\nu_{1}+\nu_{2}}\right)\frac{\alpha\beta-\gamma}{1-\beta^{2}}\right)\nn\\
\end{eqnarray}

Consider the specific case of $\nu_1=1$ and $\nu_2=1/3$ which was previously discussed in the context of $\Delta^{0}_{i,S_{1}}$, where a \tdos enhancement was observed on the $\nu_1=1$  edge when $\alpha=0$. For this case, with $\alpha = 0$, we obtain $d^{S_{1}} \simeq 2 -2\gamma (\beta+\sqrt{3})/(1-\beta^2)$ using the above equation, which implies that even for small $\beta$ and $\gamma$,  $d^{S_{1}}>1$, implying a simultaneous \tdos enhancement and stability. This fact is demonstrated clearly in the second and the third plot in Fig.~\ref{fig:S1 FP plots-pert}(a). Furthermore, the third plot in Fig.~\ref{fig:S1 FP plots-pert}(b) shows how the region of simultaneous \tdos enhancement and stability shrinks as we turn on a small but finite $\alpha$. This study established the fact that breaking of layer symmetry by taking $\nu_1 \neq \nu_2$ may lead to \tdos enhancement in one of the two layers while ensuring stability of the fixed point as was argued in the Introduction. For a finite $\alpha$ also, we do find simultaneous \tdos enhancement and stability provided we proportionately increase the strength of the other interactions,  but this is harder to see from the analytic expressions. Hence we have performed a numerical analysis to demonstrate that it is indeed possible, which is depicted in Fig.~\ref{fig:S1 FP plots}(b).

Now, consider the case when $\nu_{1} = \nu_{2} = 1$, so that $\Delta^{0}_{1} = \Delta^{0}_{2} = \Delta^{0}_{S_{1}}$. When $(\beta = \gamma = 0)$, $d^{S_{1}} = \Delta^{0}$, which is due to the fact that the ground state of the system can be written as the direct product of the ground states of individual \qh layers. In the presence of interlayer interaction $(\beta, \gamma)$, $\Delta^{0}_{S_{1}}$ and $d^{S_{1}}$ both modify themselves, and  $d^{S_{1}}$ acquires an additional contribution such that $d^{S_{1}} = \Delta^{0}_{S_{1}} + (\alpha\beta - \gamma)/(1-\beta^{2})$, which is due to the fact  that the ground states of the two \qh layers are now entangled in the presence of nontrivial $(\beta,\gamma)$. Also note that in the presence of only copropagating edge interaction $\beta$ (with $\alpha = \gamma = 0$), the ground state of the two \qh layers is still entangled, but the power laws are not modified, and we have $d^{S_{1}} = \Delta^{0}_{S_{1}}$. We do not have simultaneous \tdos enhancement and stability as expected owing to perfect layer symmetry even in the presence of $\beta,\gamma$ interaction as shown in Fig.~\ref{fig:S1 FP plots-pert}(c). We can break the layer symmetry by taking the interaction parameter $\alpha$ in the two \qh layers to be asymmetric. The analytic expressions of $\Delta^{0}_{S_{1}}$ and $d^{S_{1}}$ for the case of asymmetric $\alpha$ in the two layers are too cumbersome to be included in this paper; hence we have performed a  numerical analysis corresponding to this case and shown that the asymmetry in $\alpha$ can indeed result in simultaneous  enhancement of \tdos and stability though it requires the presence of strong interaction. The result of our numerical analysis is presented in Fig.~\ref{fig:S1 FP plots}(a). 

\section{Simultaneous \tdos enhancement and stability of $S_{2}$ fixed point}

For the strongly  coupled $S_{2}$ fixed point, the \tdos exponent for the outgoing edge of the $i$th \qh layer is denoted by $\Delta^{0}_{i,S_{2}}$ and has a  lengthy analytic expression; hence we first focus on performing an expansion of $\Delta^{0}_{i,S_{2}}$ to leading orders in $\alpha, \gamma$, which is given by

\begin{widetext}
\begin{eqnarray}
\Delta^{0}_{1,S_{2}} &\simeq& \frac{1}{\nu_{1}} \left( 1 + \frac{(\nu_{1} - \nu_{2})(\alpha - \beta\gamma)}{(1-\beta^{2})(\nu_{1} + \nu_{2})} 
 + \frac{2\sqrt{\nu_{1} \nu_{2}} (\gamma - \beta\alpha)}{(1-\beta^{2})(\nu_{1} + \nu_{2})} \right) \\
\label{Eq:pert-1}
\Delta^{0}_{2,S_{2}} &\simeq& \frac{1}{\nu_{2}} \left( 1 - \frac{(\nu_{1} - \nu_{2})(\alpha - \beta\gamma)}{(1-\beta^{2})(\nu_{1} + \nu_{2})} 
+ \frac{2\sqrt{\nu_{1} \nu_{2}} (\gamma - \beta \alpha)}{(1-\beta^{2})(\nu_{1} + \nu_{2})} \right)
\label{pertur-2}
\end{eqnarray}
\end{widetext}

\begin{figure*}[t]
\centering
\includegraphics[scale = 0.35]{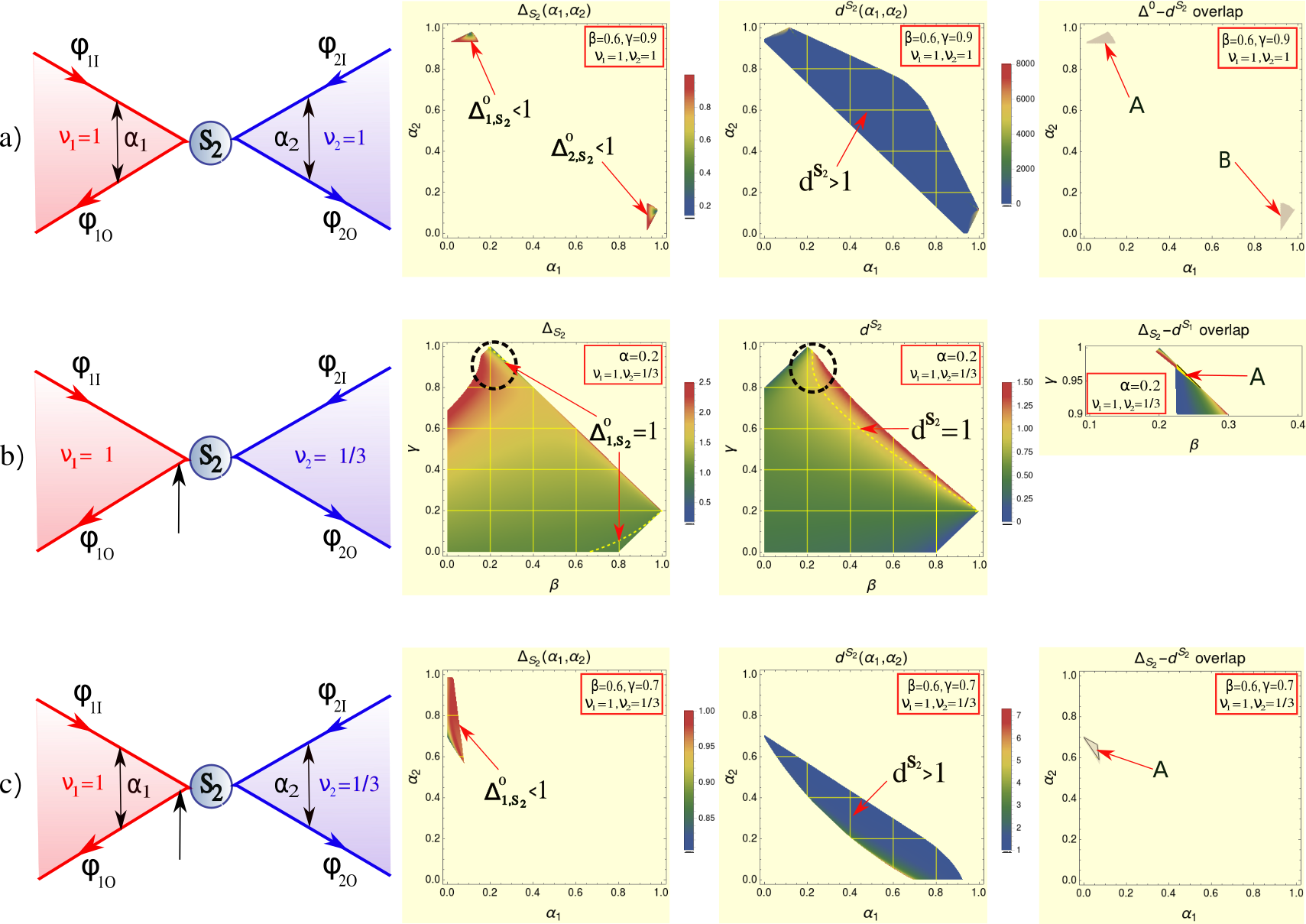}
\caption{The schematic pictures on the left show the unfolded version of the $S_{2}$ junction fixed point of bilayer \qh states exposed to a uniform magnetic field. (a) shows a junction of $\nu_{1} = 1$ and $\nu_{2} = 1$ tuned to the $S_{2}$ fixed point with asymmetric $\alpha$ in the two layers. The three density plots in (a) correspond to the region of  $\Delta^{0}_{i,S_{2}} < 1$ and $d^{S_{2}} >1$ and the region of simultaneous \tdos enhancement and stability as we move from left to right in the row for $\beta = 0.6$ and $\gamma = 0.9$. Region A (B) in the third plot shows interaction parameters for which the \tdos for the $\nu_{1}$ ($\nu_{2}$) QH edge is enhanced and the junction is simultaneously stable. (b) shows a junction of $\nu_{1} = 1$ and $\nu_{2} = 1/3$ tuned to the $S_{2}$ fixed point, in the presence of symmetric intralayer interactions. The first two density plots in (b) correspond to $\Delta^{0}_{S_{2}}$ and $d^{S_{2}}$ for $\alpha = 0.2$ as we move from left to right. The last plot in the row shows the region of simultaneous \tdos enhancement, which is marked as region A. (c) shows a junction of $\nu_{1} = 1$ and $\nu_{2} = 1/3$ tuned to the $S_{2}$ fixed point, in the presence of asymmetric $\alpha$ interactions. The first two density plots in (c) correspond to $\Delta^{0}_{S_{2}}$ and $d^{S_{2}}$ for $\beta = 0.6$ and $\gamma = 0.7$. The last plot in the row shows the region of simultaneous \tdos enhancement and stability, which is again  marked as A.
}
\label{fig:S2 FP plots_pert}
\end{figure*}

Note that in the weak $(\alpha, \gamma)$ limit, both $\Delta^{0}_{1,S_{2}} $ and $\Delta^{0}_{2,S_{2}}$  have a term which is proportional to $\nu_{1} - \nu_{2}$ but with opposite sign. This implies that if $\nu_1 > \nu_2$, then the contribution from this term will tend to suppress \tdos on the $\nu_1$ edge while it will enhance it on the $\nu_2$ edge. Now, we consider the specific case of $\nu_1=1$ and $\nu_2=1/3$ which was discussed earlier in the context of the $S_{1}$ fixed point. Naively, one would expect that it is more likely to obtain an \tdos enhancement in the $\nu_{1} = 1$ edge as compared with $\nu_{2} = 1/3$, because  the $\nu_2=1/3$ edge suffers from a strong suppression arising from the overall factor of $1/\nu$ in the expression for $\Delta^{0}_{2,S_{2}}$. Hence we focus on \tdos enhancement in the $\nu_{1} = 1$ \qh layer as it will have a higher likelihood of having simultaneous stability. Substituting $\nu_1=1$ and $\nu_2=1/3$ in the expression of  $\Delta^{0}_{1,S_{2}}$ given above, we get $\Delta^{0}_{1,S_{2}}\simeq 1+ (1/2) (\alpha - \beta\gamma)/(1-\beta^{2})+ (\sqrt{3}/2) (\gamma - \beta \alpha)(1-\beta^{2})$, which implies that if the second and the third terms in this expression turn out to be negative, then enhancement of \tdos will be possible. This would require that $\alpha < \beta\gamma$ and $\gamma < \beta \alpha$ simultaneously, which is impossible because the  interaction parameters are bounded between 0 and 1. Hence we need to look for a possibility where the sum of the two terms is negative, which implies $(\alpha - \beta\gamma)+ \sqrt{3} (\gamma - \beta \alpha)<0$. Now if we take an extreme limit of $\beta$, i.e., $\beta=1-\epsilon$, where $\epsilon$ is a small number which is of the order of $\alpha,\gamma$, or smaller and $\alpha=\gamma +\delta$ where $\delta<<\alpha,\gamma$ then the inequality reduces to $\delta (1-\sqrt{3})<0$ to leading order in all the small parameters hence resulting in \tdos enhancement. However, one should note that the  $\beta \rightarrow 1$ or equivalently the $\epsilon  \rightarrow 0$ limit of Eq.~(\ref{pertur-2}) is problematic as it is itself a perturbative result and hence we must conform it using exact numerical values. For example, for $\delta = 0.006$, $\epsilon = 0.01$, and $\gamma = 0.1$, we see \tdos enhancement on the $\nu_{1} = 1$ \qh edge, while  for $\delta = 0.00996$, $\epsilon = 0.01$, and $\gamma = 0.1$, we see \tdos enhancement on the $\nu_{1} = 1/3$ \qh edge. A numerical analysis of possible \tdos enhancement is explored in Fig.~\ref{fig:S2 FP plots_pert}(b), where we find that both for large values of $\beta (\gamma)$ and small values of $\gamma (\beta)$,  \tdos enhancement exists for $\alpha=0.2$.  

This observation of enhancement for the case of $\nu_1=1$ and $\nu_2=1/3$ is indeed very interesting when we see it in the light of  Ref.~[\onlinecite{SDas2009}], which reported \tdos enhancement for a junction of  three \lut wires in the weak repulsive interelectron interaction limit. Ref.~[\onlinecite{SDas2009}] shows a correlation between the Andreev-reflection-like process leading to hole current \cite{chamon2003} bouncing off the \lut junction and the \tdos enhancement at the junction. Note that even in our setup (which is analogous to a junction of two nonchiral  \lut wires), hole current is generated on the $\nu=1/3$ edge~\cite{Nancy, Fradkin}  for the junction of  $\nu_1=1$ and $\nu_2=1/3$. This can be seen from the expression of the field splitting matrix given in Eq.~(\ref{Eq:S2 FP}), where one of  its diagonal elements turns negative for the choice of  $\nu_1=1$ and $\nu_2=1/3$. Hence one would have naively expected that we should observe an enhancement only on the $\nu_2=1/3$ edge, but on the contrary we observe that the enhancement is happening on both the $\nu_1=1$  and the $\nu_1=1/3$ edges. We conclude that Andreev-reflection-like processes do not necessarily lead to \tdos enhancement in general.


For the $S_{2}$ fixed point, the intralayer single-quasiparticle backscattering operator represents the most relevant perturbation, and the junction is stable when $d_{11}^{O/I}, d_{22}^{O/I} > 1$. The scaling dimension is studied mostly numerically as its exact expression is too lengthy. We start by analyzing the weak $(\alpha,\gamma)$ limit by carrying out a leading order expansion of $d^{O/I}_{ii}$ in these parameters which is given by 
 
\begin{eqnarray}
d^{O/I}_{11} &\simeq& \frac{2\nu_{1}\nu_{2}}{\nu_{1} + \nu_{2}}\left[ 1+ \frac{\beta \gamma-\alpha}{1-\beta^{2}} +
\left(\frac{2\sqrt{\nu_{1}\nu_{2}}}{\nu_{1} + \nu_{2}}\right)\frac{\gamma - \beta \alpha}{1-\beta^{2}} \right]\nonumber\\
\label{Eq: back-scattering SD}
\end{eqnarray}

Note that $d^{O/I}_{11}$ is symmetric under $\nu_{1} \leftrightarrow \nu_{2}$, and thus $d^{O/I}_{11} = d^{O/I}_{22}$ as expected.  Let $d^{O/I}_{11} = d^{O/I}_{22} = d^{S_{2}}$. In the limit $\beta\gamma > \alpha$, the second term in Eq.~(\ref{Eq: back-scattering SD}) dominates over the third term, and $d^{S_{2}}$ tends towards the region where the strongly coupled $S_{2}$ fixed point is stable. Now, consider the specific case of $\nu_1=1$ and $\nu_2=1/3$. $d^{S_{2}}$ can be written as $d^{S_{2}}=1/2 + \zeta$, where $\zeta$ is a function of $\alpha,\beta,\gamma$ and is of the same order as them in the $\alpha,\beta,\gamma<<1$ limit, which implies that  the $S_{2}$ fixed point is an unstable fixed point in this limit. Above we have noted that \tdos enhancement is possible for large values of some interaction parameters [see Fig.~\ref{fig:S2 FP plots_pert}(b)] for the $\nu_1=1$ edge, and hence we would like to check whether $S_2$ can be simultaneously stable in this parameter regime; however, this analysis is too complicated to be pursued analytically owing to lengthy expressions, and hence we perform a numerical analysis. The result of our analysis is presented in last two plots in Fig.~\ref{fig:S2 FP plots_pert}(b), where we have shown the existence of a small but finite overlap region between stability and \tdos enhancement in the strong $\gamma$ limit. It is not surprising that unlike the $S_1$ fixed point, the region of simultaneous \tdos enhancement and stability of the junction for the $S_2$ fixed point always lies in the strong $\gamma$ limit. This arises from the fact that the disconnected fixed point ($S_1$) is a stable fixed point while the connected fixed point ($S_2$) is an unstable fixed point in the presence of $\alpha$ alone (i.e., $\beta=0$ and $\gamma=0$). Hence it requires large values of $\beta$ or $\gamma$ or both to stabilize the $S_2$ fixed point. Also, note that for $\nu_{1} =1$ and $\nu_{2} = 1/3$, we have $d^{O,I}_{11} = d^{O,I}_{22} = d^{O,I}_{12} = d^{O/I}_{21}$. This is due to symmetry in the current splitting matrix for the $S_{2}$ fixed point, i.e., transmission = reflectance = 1/2, when current is excited from the $\nu_{1} = 1$ side.

For $\nu_{1} = \nu_{2} = \nu$, the scaling dimensions of the tunneling operators are zero, i.e., $d^{S_{2}}_{12} = d^{S_{2}}_{21} = d^{S_{2}}_{II} = d^{S_{2}}_{OO} = 0$ as expected. Also, $\Delta^{0}_{1,S_{2}} = \Delta^{0}_{2,S_{2}} $. The exact expression for the scaling dimension of the backscattering operator and the \tdos exponent $\Delta^{0}_{i,S_{2}}$ is given by

\begin{eqnarray}
d^{S_{2}} &=& \nu \sqrt{\frac{1 - \alpha - \beta + \gamma}{1 + \alpha - \beta - \gamma}}
\label{Eq:same_nu_ds2}
\end{eqnarray}
\begin{eqnarray}
\Delta_{i,S_{2}}^{0} &=& \frac{1}{2\nu_{i}} \left( \sqrt{\frac{1 + \gamma -\beta -\alpha}{1-\gamma - \beta + \alpha}} \, + 
\sqrt{\frac{1+\alpha+\beta + \gamma}{1-\alpha + \beta -\gamma}} \right) \nn\\
\label{Eq:symmetry_S1_S2}
\end{eqnarray}

Expanding Eqs.~(\ref{Eq:same_nu_ds2}) and (\ref{Eq:symmetry_S1_S2}) in the weak $\alpha,\gamma$ limit, we get

\begin{eqnarray}
d^{S_{2}} &\simeq& \nu \left(1 + \frac{\gamma - \alpha}{1-\beta^{2}}\right) \nonumber\\
\Delta^{0}_{i,S_{2}}&\simeq& \frac{1}{\nu_{i}}\left( 1 + \frac{\gamma - \beta\alpha}{1-\beta^{2}}\right) 
\label{Eq:S2_TDOS_ds2_expansion}
\end{eqnarray}

Note that for $\nu_{1} = \nu_{2} = 1$ there exists a symmetry between the scaling dimension of the electron tunneling operator,  $d^{S_{1}}$, of the $S_{1}$ fixed point and the scaling dimension  of the electron backscattering operator, $d^{S_{2}}$, of the $S_{2}$ fixed point in the $\alpha \leftrightarrow \gamma$ exchange, such that $d^{S_{1}}(\alpha,\beta,\gamma) = d^{S_{2}}(\gamma,\beta,\alpha)$. Also, we have $\Delta^{0}_{i,S_{1}}(\alpha,\beta,\gamma) = \Delta^{0}_{i,S_{2}}(\gamma, \beta, \alpha)$. From Eq.~(\ref{Eq:S2_TDOS_ds2_expansion}), we note that in the weak $\alpha,\gamma$ limit, the junction becomes stable in the $\gamma > \alpha$ region and the \tdos shows that enhancement appears in the $\beta\alpha > \gamma$ region, which is impossible to satisfy simultaneously. Similar to the $S_{1}$ fixed point, we do not expect to see simultaneous \tdos enhancement and stability of the junction in this case, as the role of $\alpha$ and $\gamma$ gets exchanged but the region of $d^{S_{2}} > 1$ and $\Delta^{0}_{i,S_{2}} < 1$ still remains mutually exclusive. 
 
Now, we break the parity symmetry or layer symmetry of the junction by introducing asymmetric $\alpha$ in the two \qh layers as we did for the $S_{1}$ fixed point to investigate the possibility of having $d^{S_{2}} > 1$ and $\Delta^{0}_{S_{2}}<1$ simultaneously. We run a numerical search to check the possibility of simultaneous stability and \tdos enhancement in the presence of asymmetric $\alpha$ in the case of both $\nu_1=1, \nu_2=1$ and $\nu_1=1, \nu_2=1/3$, and the results of our findings are given in 
Figs.~\ref{fig:S2 FP plots_pert}(a) and \ref{fig:S2 FP plots_pert}(c). In both cases we again find the region of simultaneous stability and \tdos enhancement, but it exists only in the strong interaction limit.

\section{Discussion and conclusions}
Both the bulk and boundary of an isolated \lut wire show suppression of \tdos for  \lut parameter $K<1$, i.e., the repulsive interelectron interaction limit \cite{Fisher_Glazman}. The minimal modifications which could be added to the \lut  model such that it leads to a deviation for the standard paradigm of \tdos suppression are (i) formation of a junction of multiple \luts and  (ii) switching on nonlocal density-density interaction in addition to the local ones. Introducing exotic quantum impurity into the \lut \cite{Latief} could also lead to \tdos enhancement, but such a scenario is not the focus of this paper. The junction of \lut  wires is a well-studied subject both theoretically and experimentally, but the physical setting for motivating a nonlocal density-density interaction is not obvious. This leads us to consider the bilayer quantum Hall system, which can naturally host such a model.  In particular, a bilayer quantum Hall line junction \cite{Barkeshi} could be a possibility which allows all the four edge states participating at the junction (two from the top layer and two from the bottom layer) to come in the close vicinity of each other hence leading to mutual interactions between them. Such a system has been in discussion recently owing to the possibility of being a host to localized parafermion zero modes \cite{Barkeshi, Alicea2016, Sarma2015,Ebisu2017}. Also, there exists a long history in the experimental realization of bilayer quantum Hall systems \cite{West1990, Eisentein, Shayegan, Eisentein1992,Ensslin, Jun_Zhu}. Additionally, there has been significant experimental progress also in realizing graphene bilayer quantum Hall systems \cite{Geim,Abanin,Diankov,Kou,Maher,Kim}. This experimental progress indicates that the technology required for designing the proposed setup may not be a far-fetched one.

In general, it is difficult to find a fixed point for the \lut system which leads to \tdos enhancement at the junction of \lut s and is also stable (in the \rg sense) against perturbations that could be switched on at the junction. This is obvious as an enhancement of \tdos naturally implies the presence of relevant perturbations involving tunneling of electrons at the junction which could destabilize the junction fixed point. An important realization in this paper was the fact that simultaneous enhancement of \tdos and \rg stability of the junction fixed point is a possibility provided we break the layer symmetry either by having different filling fractions on the two layers or by choosing a different strength for the two intralayer interaction parameters ($\alpha$).

Furthermore, we would like to point out that the  occurrence of processes analogous to Andreev reflection at the junction of \lut s does not seem to provide a litmus test  for the presence of  \tdos enhancement  in general though such a connection was observed in Ref. [\onlinecite{SDas2009}] in the context of a junction of three \lut s for a repulsive interelectron interaction parameter regime. An invalidation of such an identification was demonstrated explicitly when we considered the $S_2$ fixed point between the $\nu=1$ and $\nu=1/3$ edge, which supports a process analogous to  Andreev reflection at the junction owing to the fact that the quasiparticles on the $\nu=1/3$ edge have fractional charge as opposed to the electron-like quasiparticles on the $\nu=1$ edge. Furthermore, we note that the junction of two chiral \lut s (not three) is enough to show  \tdos enhancement provided we switch on interaction (like $\beta$ and $\gamma$) in addition to the routinely considered interaction parameter $\alpha$.

Lastly, we would like to point out that here we have taken the interaction parameters to be independent of each other, which in a general QH setup need not be true. Codependencies of interaction parameters can be accounted for through a distributed circuit model introduced in Ref.~[\onlinecite{Hashisaka}] to analyze the experimental results obtained by those authors in the context of an interacting quantum Hall edge state. We apply this circuit model to our setup comprising four interacting edge states and find that the interaction
parameters indeed have strong interdependencies which cannot be ignored in general in a realistic experimental setup (see Appendix B for more details).

\begin{acknowledgments}
A.R. acknowledges University Grants Commission, India, for support in the form of a fellowship.
S.D. would  like to acknowledge the MATRICS grant ( Grant No. MTR/ 2019/001 043) from the Science and Engineering Research Board (SERB) for funding. It is a pleasure to acknowledge discussions facilitated by the International Centre for Theoretical Sciences program ``Edge Dynamics in Topological Phases" (ICTS/edytop2019/06), which initiated this work.
\end{acknowledgments}

\clearpage
\appendix

\section{Tunneling Density of states (TDOS)}
\label{TDOS}
The local electron TDOS for a chiral outgoing QH edge of a $2\times2$ QH edge junction with filling fraction $\nu_{i}$ at a point $x$ from the junction is given by
\begin{eqnarray}
\rho_{i}(E) &=& 2\pi \sum_{n} \vert\langle n | \psi^{\dagger}_{iO}(x)|0\rangle \vert^{2}\delta(E_{n} - E_{0} -E) \nonumber\\
&=& \int^{\infty}_{-\infty} \langle 0\vert \psi_{iO}(x,t)\psi^{\dagger}_{iO}(x,0)\vert0\rangle e^{-iEt} dt
\label{Eq: TDOS_eq1}
\end{eqnarray}

The fermionic field $\psi_{iI/O}$ denotes the incoming/outgoing chiral edge with filling fraction $\nu_{i}$ and can be expressed in terms of the bosonic field $\phi_{iI/O}$ as $\psi_{iI/O} \sim F_{i}e^{\iota \phi_{iI/O}/\nu_{i}}$, where $F_{i}$ is the Klein factor. Then the TDOS is given by

\begin{equation}
\rho_{i}(E) \sim \int_{-\infty}^{\infty} dt \langle 0\vert e^{i\frac{\phi_{iO}(x,t)}{\nu_{i}}}e^{-i\frac{\phi_{iO}(x,0)}{\nu_{i}}}\vert0 \rangle e^{-iEt} \\
\end{equation}

Let $\bar{\phi}_{O} = \left( \bar{\phi}_{1O}, \bar{\phi}_{2O}\right)$ and $\bar{\phi}_{I} = \left( \bar{\phi}_{1I}, \bar{\phi}_{2I}\right)$. The free Bogoliubov fields $\tilde{\phi}$ are related to the interacting $\bar{\phi}$ by the $X$ matrix, which can now be decomposed as follows:
\begin{equation}
\begin{pmatrix}
\bar{\phi}_{O} \\
\bar{\phi}_{I}
\end{pmatrix}_{(x,t)} = \begin{pmatrix}
                X_{1} & X_{2} \\
                X_{3} & X_{4}
                \end{pmatrix} \begin{pmatrix}
                              \tilde{\phi}_{O} \\
                              \tilde{\phi}_{I}
                              \end{pmatrix}_{(x,t)}
\end{equation}

The QPC of the $2\times 2$ QH edge system can be accounted for by a current splitting matrix at the junction which relates the incoming interacting bosonic fields to the outgoing interacting bosonic fields, such that 

\begin{equation}
\begin{pmatrix}
\phi_{1O} \\
\phi_{2O}
\end{pmatrix}_{(x=0)} = S \begin{pmatrix}
                              \phi_{1I} \\
                              \phi_{2I}
                              \end{pmatrix}_{(x=0)}
\end{equation} 
where $S$ denotes the current splitting matrix given by the two possible fixed points, namely, the $S_{1}$ and $S_{2}$ fixed points.
\begin{equation}
\begin{pmatrix}
\bar{\phi}_{1O} \\
\bar{\phi}_{2O}
\end{pmatrix}_{(x=0)} = M^{-1} S M \begin{pmatrix}
                              	   \bar{\phi}_{1I} \\
                              	   \bar{\phi}_{2I}
                              	   \end{pmatrix}_{(x=0)} = \bar{S}\begin{pmatrix}
                              	   							\bar{\phi}_{1L} \\
                              	   							\bar{\phi}_{2L}
                              	   								\end{pmatrix}_{(x=0)} \\
\end{equation}
where $M_{ij} = \sqrt{\nu_{i}}\delta_{ij}$. Then the real interacting bosonic fields $\phi_{O}$ and $\phi_{I}$ can be expressed only in terms of the left-moving Bogoliubov field $\tilde{\phi}_{I}$ ( which are independent of each other) as follows:
\begin{equation}
\phi_{O}(x,t) = M\left[T_{1} \tilde{\phi}_{I}(-x,t) + T_{2}\tilde{\phi}_{I}(x,t)\right],   
\end{equation}
\begin{equation}
\phi_{I}(x,t) = M\left[T_{3} \tilde{\phi}_{I}(-x,t) + T_{4}\tilde{\phi}_{I}(x,t)\right],    
\end{equation}
where
\begin{eqnarray}
T_{1} &=& X_{1}\left(X_{1} - \bar{S}X_{3}\right)^{-1}\left(\bar{S}X_{4}-X_{2}\right), \\
T_{2} &=& X_{2}, \\
T_{3} &=& X_{3}\left(X_{1} - \bar{S}X_{3}\right)^{-1}\left(\bar{S}X_{4}-X_{2}\right), \\
T_{4} &=& X_{4}.
\end{eqnarray}
Then
\begin{eqnarray}
\langle 0\vert \psi_{iO}(x,t)\psi^{\dagger}_{iO}(x,0)\vert0\rangle &\sim& \prod_{j=1}^{2}\left( \frac{i\alpha}{-\tilde{v}_{j}t + i\alpha}\right)^{\gamma_{ij}}  \nonumber\\
&& \times\left(\frac{(i\alpha)^{2}-4x^{2}}{(i\alpha-\tilde{v}_{j}t)^{2}-4x^{2}}\right)^{\zeta_{ij}}, \nonumber\\
\label{Eq:Gnrl_TDOS_correlator}
\end{eqnarray}
where $\alpha$ is the short-distance cutoff, $\gamma_{ij}=\frac{\left[T_{1}\right]^{2}_{ij} + \left[T_{2}\right]^{2}_{ij}}{\nu_{i}}$, and $\zeta_{ij} = \frac{\left[T_{1}\right]_{ij}\left[T_{2}\right]_{ij}}{\nu_{i}}$. Now we calculate the TDOS in two limits, namely, first at the junction with $x\longrightarrow 0$, in which case Eq. (\ref{Eq:Gnrl_TDOS_correlator}) becomes
\begin{equation}
\langle 0\vert \psi_{iO}(x,t)\psi^{\dagger}_{iO}(x,0)\vert0\rangle \sim \prod_{j=1}^{2}\left( \frac{i\alpha}{-\tilde{v}_{j}t + i\alpha}\right)^{\gamma_{ij} + 2\zeta_{ij}}, 
\label{Eq:TDOS_correlator_x=0}  
\end{equation}
and the other far from the junction with $x \longrightarrow \infty$, in which case Eq. (\ref{Eq:Gnrl_TDOS_correlator}) becomes
\begin{equation}
\langle 0\vert \psi_{iO}(x,t)\psi^{\dagger}_{iO}(x,0)\vert0\rangle \sim \prod_{j=1}^{2}\left( \frac{i\alpha}{-\tilde{v}_{j}t + i\alpha}\right)^{\gamma_{ij}}.  
\label{Eq:TDOS_correlator_x_infinity}
\end{equation}

Now from Eqs.~(\ref{Eq: TDOS_eq1}), (\ref{Eq:TDOS_correlator_x=0}), and (\ref{Eq:TDOS_correlator_x_infinity}), we have the TDOS integral in the two limits of the form
\begin{equation}
\begin{aligned}
\int_{-\infty}^{\infty} \prod_{j=1}^{2}\left( \frac{i\alpha}{-\tilde{v}_{j}t + i\alpha}\right)^{\Delta_{ij}} e^{-iEt}dt ~\propto {} & ~E^{\left(\Delta_{i} -1\right)}\\
\end{aligned},
\end{equation}
where the TDOS exponent $\Delta_{i}$ is given by
\begin{equation}
\Delta^{0}_{i} = \frac{1}{\nu_{i}}\sum_{j=1}^{2}\left(\left[T_{1}\right]_{ij}+\left[T_{2}\right]_{ij}\right)^{2}     
\end{equation}
and, far from the junction,
\begin{equation}
\Delta^{\infty}_{i} = \frac{1}{\nu_{i}}\sum_{j=1}^{2}\left(\left[T_{1}\right]^{2}_{ij}+\left[T_{2}\right]^{2}_{ij}\right).     
\end{equation}

\section{Inter-dependency of Interaction Parameters}

Let us consider the simplest possible case of two quantum Hall (QH) systems with filling fraction $\nu_{1} = \nu_{2} = 1$ in bilayer stacking, with two incoming and two outgoing edges. Here, we use a simple approach to account for the effect of Coulomb interactions between the edges in terms of a distributed circuit model \cite{Hashisaka}. The dynamics of edge plasmons traveling along a single edge channel is modeled through the distributed electrochemical capacitance per unit length between the channel and the ground (denoted by $C_{ch}$-channel capacitance). The interaction between the two different channels is modeled with distributed elements, which is expressed by the interedge capacitance.

Interedge capacitance per unit length between $\phi_{iI}$ and $\phi_{iO}$ of the same QH layer is given by $C_{\alpha}$, between $\phi_{iI}$ and $\phi_{jI}$ of different QH layers ($i\neq j$) is given by $C_{\beta}$, and between $\phi_{iI}$ and $\phi_{jO}$ of different QH layers is given by $C_{\gamma}$. Channel capacitance per unit length  $C_{ch}^{(i)}=C_{ch}$ for all the edges ($ i = \lbrace 1,4\rbrace$) is taken to be the same (as the Fermi velocity for each edge plasmon is taken to be the same \cite{Hashisaka}). Let $\rho_{iO/I}(x,t)$, $V_{iO/I}(x,t)$, and $I_{iO/I}(x,t)$ be the excess charge density, potential, and current flowing through the out/in edge channel of the $i$th QH layer, respectively, at position $x$ and time $t$. The relation between the current $I_{i}(x,t) = (I_{1O}, I_{2O}, I_{1I}, I_{2I})^{T}$ and potential $V_{i}(x,t) = (V_{1O}, V_{2O}, V_{1I}, V_{2I})^{T}$ is given by, $I_{iO}(x,t) = \sigma_{xy}^{(i)}V_{iO}(x,t)$ and $I_{iI}(x,t) = -\sigma_{xy}^{(i)}V_{iI}(x,t)$.

The excess charge density $\rho_{iI/O}$ is related to the potential through the matrix $C_{T}$ given by

\begin{equation}
\begin{pmatrix}
\rho_{1O}\\
\rho_{2O}\\
\rho_{1I}\\
\rho_{2I}
\end{pmatrix}_{(x,t)} = \begin{pmatrix}
						C_{p} & -C_{\beta} & -C_{\alpha} & -C_{\gamma} \\
						-C_{\beta} & C_{p} & -C_{\gamma} & -C_{\alpha} \\
						-C_{\alpha} & -C_{\gamma} & C_{p} & -C_{\beta} \\
						-C_{\gamma} & -C_{\alpha} & -C_{\beta} & C_{p}
						\end{pmatrix}   \begin{pmatrix}
										V_{1O}\\
										V_{2O}\\
										V_{1I}\\
										V_{2I}
										\end{pmatrix}_{(x,t)}
\label{Eq:Capacitor_charge_relation}	
\end{equation}

which in compacted form can be written as $\rho(x,t) = C_{T}V(x,t)$. $C_{p}$ is given by $C_{p} = C_{ch}+C_{\alpha} + C_{\beta} + C_{\gamma}$. The Heisenberg equation of motion for the coupled system [Eq. 8 of the main text] is given by
\begin{equation}
\frac{d}{dt}\phi_{a}(x,t) = -v_{f}\epsilon_{a}\sum_{\alpha=1}^{4} K_{a\alpha} \frac{d}{dx}\phi_{\alpha}(x,t). 
\end{equation}
Since $\rho_{iI/O} = \pm \frac{d}{dx}\phi_{iI/O}$ and $I_{iI/O} = \mp \frac{d}{dt} \phi_{iI/O}$, we have
\begin{eqnarray}
\frac{d}{dt}I_{a}(x,t) &=& -v_{f}\sum_{\alpha=1}^{4} K_{a\alpha}\frac{d}{dx}I_{\alpha}(x,t), \nonumber\\
\frac{d}{dt}I(x,t)  &=& -U\frac{d}{dx}I(x,t),
\end{eqnarray}

where $U$ is a $4\times 4$ matrix given by
\begin{equation}
U = vf\begin{pmatrix}
	1 & \beta & \alpha & \gamma \\
	\beta & 1 & \gamma & \alpha \\
	-\alpha &  \gamma & -1 & -\beta \\
	-\gamma & -\alpha & -\beta & -1
	\end{pmatrix}.
\end{equation}

Now, using the continuity equation $\partial_{t}\rho_{i,I/O} + \partial_{x}I_{i,I/O} = 0$, Eq.~(\ref{Eq:Capacitor_charge_relation}), and the relation $I_{iO/I}(x,t) = \pm\sigma^{(i)}_{xy}V_{iO/I}(x,t)$, we get
\begin{equation}
U = \sigma_{xy} C_{T}^{-1},
\end{equation}
where $\sigma_{xy}$ is a diagonal 4$\times$4 matrix with $(\sigma^{(1)}_{xy}, \sigma^{(2)}_{xy}, -\sigma^{(1)}_{xy}, -\sigma^{(2)}_{xy})$ being the diagonal elements. We can now express interaction parameters in terms of capacitance as follows:
\begin{eqnarray}
v_{f} &=& \frac{1}{C_{ch}} + \frac{1}{C_{ch} + 2\left(C_{\alpha} + C_{\beta}\right)} + \frac{1}{C_{ch} + 2\left(C_{\alpha} + C_{\gamma}\right)} \nonumber\\
&& + \frac{1}{C_{ch} + 2\left(C_{\beta} + C_{\gamma}\right)} \nonumber\\
\alpha &=& \frac{\frac{1}{C_{ch}} - \frac{1}{C_{ch} + 2\left(C_{\alpha} + C_{\beta}\right)} - \frac{1}{C_{ch} + 2\left(C_{\alpha} + C_{\gamma}\right)} + \frac{1}{C_{ch} + 2\left(C_{\beta} + C_{\gamma}\right)}}{\frac{1}{C_{ch}} + \frac{1}{C_{ch} + 2\left(C_{\alpha} + C_{\beta}\right)} + \frac{1}{C_{ch} + 2\left(C_{\alpha} + C_{\gamma}\right)} + \frac{1}{C_{ch} + 2\left(C_{\beta} + C_{\gamma}\right)}  } \nonumber\\
\beta &=& \frac{\frac{1}{C_{ch}} - \frac{1}{C_{ch} + 2\left(C_{\alpha} + C_{\beta}\right)} + \frac{1}{C_{ch} + 2\left(C_{\alpha} + C_{\gamma}\right)} - \frac{1}{C_{ch} + 2\left(C_{\beta} + C_{\gamma}\right)}}{\frac{1}{C_{ch}} + \frac{1}{C_{ch} + 2\left(C_{\alpha} + C_{\beta}\right)} + \frac{1}{C_{ch} + 2\left(C_{\alpha} + C_{\gamma}\right)} + \frac{1}{C_{ch} + 2\left(C_{\beta} + C_{\gamma}\right)}  } \nonumber\\
\gamma &=& \frac{\frac{1}{C_{ch}} + \frac{1}{C_{ch} + 2\left(C_{\alpha} + C_{\beta}\right)} - \frac{1}{C_{ch} + 2\left(C_{\alpha} + C_{\gamma}\right)}- \frac{1}{C_{ch} + 2\left(C_{\beta} + C_{\gamma}\right)}}{\frac{1}{C_{ch}} + \frac{1}{C_{ch} + 2\left(C_{\alpha} + C_{\beta}\right)} + \frac{1}{C_{ch} + 2\left(C_{\alpha} + C_{\gamma}\right)} + \frac{1}{C_{ch} + 2\left(C_{\beta} + C_{\gamma}\right)}  } \nonumber\\
\label{Eq:int_parameter_dependency}
\end{eqnarray}

As can be seen from the above equation~(\ref{Eq:int_parameter_dependency}), interactions between the edges, in general, cannot be treated as independent parameters in a realistic situation.

\end{document}